\documentclass[12pt]{article}  
\usepackage{amsmath,amssymb,bm}
\usepackage{epsfig}\usepackage{graphicx}
\usepackage{amsfonts}
\usepackage{epstopdf}
\usepackage[usenames,dvipsnames]{color} 
\usepackage{float}

\newcommand{\beqa}{\begin{eqnarray}}
\newcommand{\eeqa}{\end{eqnarray}}
\newcommand{\bd}[1]{ \mbox{\boldmath $#1$}}

\begin{document}
\def\ii{\'\i}

\title{Semi-microscopic construction of multi-$\alpha$
cluster spaces}

\author{
J. R. M. Berriel-Aguayo$^1$
and 
P. O. Hess$^{1,2}$ \\
{\small\it
$^1$ Instituto de Ciencias Nucleares, Universidad Nacional 
Aut\'onoma de M\'exico,}\\
{\small\it
Circuito Exterior, C.U., 
A.P. 70-543, 04510 M\'exico D.F., Mexico}\\
{\small\it
$^2$ Frankfurt Institute for Advanced Studies, Johann Wolfgang 
Goethe Universit\"at,}\\
{\small\it
Ruth-Moufang-Str. 1, 60438 Frankfurt am Main, Germany}

}

\maketitle

\abstract{
An approximate but straight forward 
projection method to molecular many $\alpha$-particle
states is proposed and the overlap to the shell model space
is determined. The resulting space is in accordance
with the shell model, but still contains states which are not
completely symmetric under permutations of the 
$\alpha$-particles, which is one reason to call the construction {\it semi-microscopic}. 
A new contribution
is the construction of the 6- and 7-$\alpha$-particle spaces.
The errors of the method propagate toward larger number of
$\alpha$-particles and larger shell excitations.
In order to show the effectiveness of the construction 
proposed,
the so obtained spaces are applied, 
within an algebraic cluster model, 
to $^{20}$Ne, $^{24}$Mg and $^{28}$Si, 
each treated
as a many-$\alpha$-particle system. 
Former results on $^{12}$C and $^{16}$O are resumed
%\PACS{
%      {21.00}{nuclear structure}   \and
%      {21.60.Cs}{shell model} \and
%			{21.60.Gx}{cluster model}
%     } % end of PACS codes
%\keywords{nuclear clusters, algebraic model, heavy nuclei}
}

\section{Introduction}
\label{intro}

Cluster physics has always enjoyed a great interest,
because it
enlightens the structure of nuclei 
and provides a manageable method to treat, in general, many 
cluster systems, avoiding the very large shell model
space.
\cite{wildermuth}. In particular, the $\alpha$-cluster
structure plays an important role in understanding the
microscopic underpinning of light and heavy nuclei. For a
recent
review on the experimental findings and theoretical
ones on the cluster structure in light nuclei, 
see \cite{freer2018} and \cite{yamada2012},
respectively. 

An essential ingredient of any cluster model is the
construction of the Hilbert space. For example,
in \cite{kato} a microscopic projection method was applied 
for the construction of up to the 5-$\alpha$ particle space, besides other
exotic cluster configurations. The method is quite involved
and requires to calculate integral kernels, which 
are, however, subject to numerical errors.

The $\alpha$-cluster structure is important in
several areas, as decay modes of light and heavy nuclei
(for example in the reaction $\alpha$ + $^{16}$O), fusion reactions in heavy stars and investigations of 
$\alpha$-particle condensates in nuclei \cite{schuck2018}.
This illustrates the necessity to obtain the Hilbert
space for any number of $\alpha$-particle systems,
without recurring to involved 
methods for the construction of these spaces.

The main objective of this contribution is to introduce
an iterative procedure to obtain, at least approximately, 
the microscopic space for any
number of $\alpha$-particle system, which at the same
time is practical and straight forward to apply. The method is
based only on algebraic manipulations and
we intend to convince 
that an approximate projection is favored over an exact one.
We will see that 
for systems of few $\alpha$-particles it works for low lying 
states and it does not suffer from involved manipulations
as in \cite{kato}. Especially, we will see that
the results in \cite{kato} also suffer from states which are
not part of the shell model space and that the origin 
lies in the calculation
of overlap matrix elements which generates numerical errors.
However,
the price to pay for the approximation we use, is the inclusion of states which
are not completely symmetric in the permutations of the
$\alpha$-particles. 

The $\alpha$-cluster spaces are by itself
useful for any theory on
many $\alpha$-cluster systems. In order to
illustrate the application of the classification we will only apply a phenomenological and algebraic cluster model, namely the so-called {\it Semimicroscopic Algebraic Cluster
Model} (SACM) \cite{cseh-letter,cseh-levai-anph}, which in 
most cases is effective in two-cluster systems, also
as an alternative to the geometric description of a
two-cluster system \cite{scheid1995,hess-1984}. Spectra,  
transition values and spectroscopic factors are 
also calculated.

In what follows, the iterative procedure will be explained,
which is applied to 3 ($^{12}$C), 4 ($^{16}$O), 
5 ($^{20}$Ne), 6 ($^{24}$Mg) and and 7 ($^{28}$Si)  
$\alpha$-particle states. 
A new contribution is the
list of 6- and 7-$\alpha$ particle states. 

The paper is structured as follows: In Section
\ref{general} the iterative procedure is explained,
whose results will be compared to 
\cite{kato} for up to 5 $\alpha$-particles, 
were we will also prove that the space of the 
5-$\alpha$ system, as determined in  \cite{kato}, 
includes states which are not
allowed by the PEP.  In Section \ref{appl} we apply the
results obtained to systems of up to 7 $\alpha$-particles. 
In Section \ref{concl} Conclusions will be drawn.
In the Appendix we construct explicitly the
5 $\alpha$-particle space of $0\hbar\omega$ and
$1\hbar\omega$.

\section{Approximate construction of the multi 
$\alpha$-particle space}
\label{general}

The objective is to construct a basis for a multi $\alpha$ 
cluster systems, whose states are classified by a ket,
depending on $SU(3)$ quantum numbers, the natural
choice for the shell model. 
The molecular basis 
has an overlap to the shell model, which is easy
to satisfy. The main difficulty is to achieve that these 
basis states also are symmetric
under the interchange of any two $\alpha$ particles. The
method is approximate, with the
goal that at least states which are low lying in energy 
are symmetric under the permutation of the $\alpha$ particles.
The result will be compared to existing constructions of
3-, 4- and 5-$\alpha$ particle systems.

The construction of the multi-$\alpha$ particle space
is divided into two steps:

\begin{itemize}

\item {\bf Step 1}: Construction of molecular 
$\alpha$-particle states. The $\alpha$-particles will be
added successively in such a manner that the symmetry between
them is at least approximately satisfied.

\item {\bf Step2}: The shell model space is explicitly
constructed, to which the molecular states are compared and
the overlap is determined. The resulting states are the
ones which are taken into account as part of the model space.
With increasing $n\hbar\omega$ shell excitations, more and
more states arise, which are not symmetric under the 
interchange of $\alpha$-particles and also the multiplicity
is larger than for the allowed ones. Fortunately, for low
$n$-values and/or large eigenvalues of the second order
$SU(3)$ Casimir operator, the space is free of
these spurious states.

\end{itemize}

Before proceeding, we have to spent some words on
the definition of clusterization,
there are two definitions: The weak
and the strong one. In the strong definition, the nuclear
subsystem has to exhibit an explicit spatial separation from
the other clusters, while in the weak definition this is not 
the case, only requiring that the cluster state has an
overlap to the state of the united nucleus. E.g.,
for a two-cluster system the overlap is 
$\langle \Psi \mid \psi_{c_1}\psi_{c_2} R\rangle$,
where $\mid \Psi\rangle$ is the state of the united nucleus,
$\psi_{C_k}$ ($k=1,2$) are the cluster states and $R$ is
the relative wave function.
For example,
the ground state of $^{16}$O is a spherical nucleus 
which has an overlap to the cluster state of
$^{12}$C+$\alpha$, thus it shows a clusterization in the
weak sense but not in the strong sense, because
$^{26}$O is spherical. Here, we
apply the definition of the weak clusterization. 

\subsection{Iterative procedure for the molecular states}
\label{iter-proc}

In the construction of the molecular states we 
proceed as follows: The relative position of $n$ 
$\alpha$-particles are traced via $(n-1)$ 
{\it Jacobi-coordinates}, where  
${\vec \lambda}_k$ ($k=1,2,...,n-1$) is the Jacobi
coordinate between the $(k+1)$'st $\alpha$-particle and
the center of mass of the subsystem of the first $k$ 
$\alpha$-particles.

Each of these $\vec{\lambda_k}$ refer to a relative motions,
which is described by an harmonic oscillator, and therefore
carry $N_{\lambda_k}$ oscillation quanta.

We impose the restriction

\beqa
N_{\lambda_1} & \ge & N_{\lambda_2} ~\ge~
...~\ge~ N_{\lambda_{n-1}}
~~~. 
\label{eq1}
\eeqa
 
With the restriction (\ref{eq1}) 
the original state and the one
where the $N_{k_1}$ 
is interchanged with $N_{k_2}$ are 
treated as equal, corresponding to 
having permuted the two $\alpha$ particles. 
Thus, the two
states are symmetric to each other:
Under the permutation of the particle $k_1$ with $k_2$,
the states change accordingly

\beqa
P_{k_1,k_2}
\mid N_1 ... N_{k_1} ... N_{k_2} ... N_{k_{n-1}}\rangle
=
\mid N_1 ... N_{k_2} ... N_{k_1} ... N_{k_{n-1}}\rangle
~~~.
\label{eq1a}
\eeqa

The $SU(3)$-irreducible representations (irreps) 
are obtained by multiplying
successively the
$SU(3)$ irreps of the $k$-$\alpha$-particle space
with $(N_{\lambda_{k+1}},0)$, imposing the condition 
(\ref{eq1}). In this manner, the state 
(\ref{eq1a}) describes approximately a
symmetric $\alpha$-particle configurations. An explicit 
symmetrization, as done in \cite{kato} or in
\cite{kramer,mosh-smirnov}, is not applied.
We also do not construct the coordinate representation
of the cluster wave function, but rather use the ket-notation.
The advantage is that we only need to demand
that the quantum numbers correspond to a state 
of a multi-$\alpha$ particle configuration, 
with an overlap to a state within the
shell model with the same quantum numbers. 
The disadvantage, though, is that we cannot
address the strong definition of clusterization. 

In this manner, we obtain iteratively the space of 3- to 
7-$\alpha$-particle states, listed in the Tables
\ref{tab1}, \ref{tab2}, \ref{tab3}, \ref{tab6alpha}
and \ref{tab7alpha}.
The procedure does not involve an
{\it explicit} 
construction of symmetric states, but only partially
by the above condition imposed. The result is 
compared to the shell model space, whose construction is
explained in the Appendix at hand of the $1\hbar\omega$
excitation of $^{20}$Ne.

{\it The advantage of the method proposed is the direct 
extension to any number of $\alpha$-particles.}

The obtained approximate $n$ $\alpha$-particle 
space are compared to the work of Horiuchi  \cite{horiuchi}
and Kat${\bar {\rm o}}$ \cite{kato}. For small $\hbar\omega$ excitations
the list is {\it identical} and for large $\hbar\omega$
excitations, some differences in the multiplicity of
{\it high lying states} (smaller eigenvalue of the 
second order Casimir operator) appear and also some
new irreps (i.e. not satisfying the symmetry property under
permutations of the $\alpha$-particles), but also with
a smaller eigenvalue of the second order Casimir operator.

\begin{center}
\begin{table}[h!]
\centering
\begin{tabular}{|c|c|}
\hline\hline
%ind&&
$n\hbar\omega$ & $\left(\lambda , \mu\right)$ \\
\hline
0 & (0,4) \\
1 & (3,3)  \\
2 & (2,4), (4,3), (6,2)  \\
3 & (3,4), (5,3), (7,2), (9,1)  \\
4 & (4,4), (6,3), (8,2), (10,1), (12,0)  \\
5 & (5,4), (7,3), (13,0)  \\
6 & (6,4), (8,3), (10,2), (12,1)  \\
\hline
 \end{tabular}
\caption{
Model space of the $^{12}$C nucleus within the SACM for up to $6\hbar\omega$ excitations. 
The content agrees 
with \cite{horiuchi}.
} 
\vspace{0.2cm}
\label{tab1}
\end{table}
\end{center}

\begin{center}
\begin{table}[h!]
\centering
\begin{tabular}{|c|c|c|}
\hline
$n \hbar \omega$ & Kat${\bar {\rm o}}$ & present comilation\\
\hline
0 & $(0,0)$ & $(0,0)$  \\ 
 &  & \\
1 & $(2,1)$ & $(2,1)$  \\
 &  & \\
2 & $(2,0) \ (3,1) \ (0,4) \ (4,2)$ & $(2,0)^{\textbf{\textcolor{red}{2}}} \ (3,1) \ (0,4) \ (4,2)$  \\
 &  & \\
 3 & $(3,0) \ (0,3) \ (2,2) \ (4,1) \ $ & $ \textit{\textcolor{red}{(0,0)}} \ \textit{\textcolor{red}{(1,1)}}^{\textbf{\textcolor{red}{2}}} \ (3,0)^{\textbf{\textcolor{red}{3}}} \ (0,3)^{\textbf{\textcolor{red}{2}}} \ $ \\
 & $ (3,3) \ (6,0) \ (5.2) \ (2,5) \ $ & $ (2,2)^{\textbf{\textcolor{red}{2}}} \ (4,1)^{\textbf{\textcolor{red}{3}}} \ \textit{\textcolor{red}{(1,4)}} \ (3,3) \   $ \\
 & $(6,3)$ & $ (6,0) \ (5,2)  \ (2,5) \ (6,3)$ \\
 &  & \\
\hline
\end{tabular}
\caption{Space of the 4-$\alpha$-particle system.
In italic
we denote the irreps in the right column which do not
appear in the left column. For $1\hbar\omega$ the $(2,0)$
irrep appears twice in the present calculation, while
in $\cite{kato}$ the multiplicity is one. At $3\hbar\omega$
excitation, besides an increase in the multiplicities also
additional irreps appear in our compilation.
\label{tab2}}
\end{table}
\end{center}

\begin{center}
\begin{table}[h!]
\centering
\begin{tabular}{|c|c|c|}
\hline
$n \hbar \omega$ & Kat${\bar {\rm o}}$ & present compilation\\
\hline
0 & $(2,0) \ (0,4) \ (4,2) \ (8,0)$ & $(2,0) \ (0,4) \ (4,2) \ (8,0)$  \\ 
 &  & \\
1 & $(3,0) \ (0,3) \ (2,2) \ (4,1) \  $ & $ \textit{\textcolor{red}{(1,1)}}^{\textbf{\textcolor{red}{2}}} \ (3,0)^{\textbf{\textcolor{red}{3}}} \ (0,3)^{\textbf{\textcolor{red}{2}}} \ (2,2)^{\textbf{\textcolor{red}{4}}}   $ \\
 & $ (1,4) \ (3,3)^2 \ (5,2)^2 \ (2,5) \  $ & $ (4,1)^{\textbf{\textcolor{red}{3}}} \ (1,4)^{\textbf{\textcolor{red}{2}}} \ (3,3)^{\textbf{\textcolor{red}{4}}} \ \textit{\textcolor{red}{(6,0)}}  $  \\
 & $ (4,4) \ (7,1) \ (6,3) \ (9,0) $ & $ (5,2)^{\textbf{\textcolor{red}{4}}} \ (2,5)  \ (4,4) \ (7,1)^{\textbf{\textcolor{red}{2}}} $ \\
 & $ \textit{\textcolor{blue}{(5,5)}}^{\textcolor{blue}{2}} \ (8,2)^{\textbf{\textcolor{blue}{2}}} \ \textit{\textcolor{blue}{(7,4)}}^{\textcolor{blue}{3}} \textit{\textcolor{blue}{(9,3)}}^{\textcolor{blue}{2}} \ $ & $(6,3) \ (9,0) \ (8,2)$ \\
 & $\textit{\textcolor{blue}{(8,5)}} \ \textit{\textcolor{blue}{(10,4)}}$ & \\
 &  & \\
2 & $(0,2)^2 \ (2,1) \ (1,3)^2 \ (4,0)^3  $ & $\textit{\textcolor{red}{(1,0)}} \ (0,2)^{\textbf{\textcolor{red}{6}}} \ (2,1)^{\textbf{\textcolor{red}{10}}} \ (1,3)^{\textbf{\textcolor{red}{11}}}   $  \\
 & $ (3,2)^2 \ (2,4)^5 \ (5,1)^3 \ (4,3)^4  $ & $ (4,0)^{\textbf{\textcolor{red}{13}}} \ (3,2)^{\textbf{\textcolor{red}{14}}} \ \textit{\textcolor{red}{(0,5)}} \ (2,4)^{\textbf{\textcolor{red}{15}}}  $ \\
 & $  (1,6)^2 \ (3,5)^2 \ (6,2)^6 \ (5,4)^5  $ & $ (5,1)^{\textbf{\textcolor{red}{12}}} \ (4,3)^{\textbf{\textcolor{red}{14}}}  \ (1,6)^{2} \ \textit{\textcolor{red}{(7,0)}}  $\\
 & $ (0,8) \ (8,1)^4  \ (4,6)^4 \ (7,3)^6 $ & $ (3,5)^{\textbf{\textcolor{red}{4}}} \ (6,2)^{\textbf{\textcolor{red}{14}}} \ (5,4)^{\textbf{\textcolor{red}{4}}} \ (0,8) $ \\
  & $ (6,5)^{\textbf{\textcolor{blue}{6}}} \ (10,0)^3 \ (9,2)^{\textbf{\textcolor{blue}{2}}} \ (8,4)^{\textbf{\textcolor{blue}{10}}} $ & $ (8,1)^{\textbf{\textcolor{red}{5}}} \ (4,6)^{\textbf{\textcolor{red}{2}}} \ (7,3)^{\textbf{\textcolor{red}{3}}} \ (6,5) $ \\
 & $ (11,1) \ \textit{\textcolor{blue}{(7,6)}}^{\textcolor{blue}{2}} \ \textit{\textcolor{blue}{(10,3)}}^{\textcolor{blue}{4}} \ \textit{\textcolor{blue}{(9,5)}}^{\textcolor{blue}{4}} $ & $(10,0)^{3} \ (9,2) \ (8,4) \ (11,1)$ \\
 & $\textit{\textcolor{blue}{(11,4)}}^{\textcolor{blue}{2}} \ \textit{\textcolor{blue}{(13,3)}}$ & \\
 &  & \\
3 & $ (0,1) \ (1,2)^3  \ (3,1)^4 \ (0,4)   $ & $ (0,1)^{\textbf{\textcolor{red}{6}}} \ \textit{\textcolor{red}{(2,0)}}^{\textbf{\textcolor{red}{11}}} \ (1,2)^{\textbf{\textcolor{red}{31}}} \ (3,1)^{\textbf{\textcolor{red}{47}}} \ $  \\
 & $ (2,3)^6 \  (5,0)^4 \ (4,2)^6 \ (1,5)^5   $ & $ (0,4)^{\textbf{\textcolor{red}{19}}} \ (2,3)^{\textbf{\textcolor{red}{60}}} \ (5,0)^{\textbf{\textcolor{red}{42}}} \ (4,2)^{\textbf{\textcolor{red}{68}}} $\\ 
 & $ (3,4)^9 \ (6,1)^6  \ (0,7)^2 \ (5,3)^{11}  $ & $ (1,5)^{\textbf{\textcolor{red}{68}}} \ (3,4)^{\textbf{\textcolor{red}{63}}} \ (6,1)^{\textbf{\textcolor{red}{48}}} \ (0,7)^{\textbf{\textcolor{red}{8}}}  $\\
 & $ (2,6)^5 \ (8,0)^3   \ (4,5)^9 \ (7,2)^{11}   $ & $ (5,3)^{\textbf{\textcolor{red}{62}}} \ (2,6)^{\textbf{\textcolor{red}{15}}} \ (8,0)^{\textbf{\textcolor{red}{12}}} \  (4,5)^{\textbf{\textcolor{red}{25}}} $\\
 & $ (1,8)^2 \ (6,4)^9 \ (3,7)^4 \ (9,1)^4  $ & $ (7,2)^{\textbf{\textcolor{red}{45}}} \ (1,8)^{\textbf{\textcolor{red}{3}}} \ (6,4)^{\textbf{\textcolor{red}{20}}} \ (3,7)^{\textbf{\textcolor{red}{5}}}  $ \\
 & $ (5,6)^6 \ (8,3)^{12} \ (2,9) \ (7,5)^{16}  $ & $ (9,1)^{\textbf{\textcolor{red}{19}}} \ (5,6)^6 \ (8,3)^{\textbf{\textcolor{red}{13}}} \ (2,9)  $ \\
 & $ (4,8) \ (11,0)^4 \ (10,2)^5 (6,7)^{\textbf{\textcolor{blue}{3}}} \ $ & $ (7,5)^{\textbf{\textcolor{red}{4}}} \ (4,8) \ (11,0)^{\textbf{\textcolor{red}{8}}} \ (10,2)^{\textbf{\textcolor{red}{4}}}  $ \\
 & $ (9,4)^{18} \ (8,6)^{\textbf{\textcolor{blue}{12}}} \ (12,1) \ (11,3)^{\textbf{\textcolor{blue}{8}}}  $ & $ (6,7) \ (9,4)^{\textbf{\textcolor{red}{2}}} \ (8,6) \ (12,1)^{\textbf{\textcolor{red}{2}}} $ \\
 & $\textit{\textcolor{blue}{(10,5)}}^{\textcolor{blue}{8}} \ \textit{\textcolor{blue}{(13,2)}} \ \textit{\textcolor{blue}{(12,4)}}^{\textcolor{blue}{4}} \ \textit{\textcolor{blue}{(14,3)}}^{\textcolor{blue}{2}}$ & $(11,3)$ \\
 &  & \\
\hline
\end{tabular}
\caption{Space of the 5-$\alpha$-particle system.
In bold-face in the left column the different multiplicity
is indicated, compared to the right column. In italic
we denote the irreps in the right column which do not
appear in the left column.
(For color online: The blue tainted irreps in
the left column refer to those
not appearing in the present compilation. 
The blue irreps in the right column appear only in
the present compilation and the red tainted multiplicities
are larger than in Kat${\bar {\rm o}}$'s compilation).
Note, as an example, the $(10,4)$ irreps on the left side,
which is not present on the right side. In the Appendix
we prove that this state has to be excluded because it
is not part of the shell model space.
}
\vspace{0.2cm}
\label{tab3}
\end{table}
\end{center}

\begin{center}
\begin{table}[h!]
\centering
\begin{tabular}{|c|c|}
\hline
$n \hbar \omega$ & $6-\alpha$ \\
\hline
0 & $(0,2) \ (1,3) \ (4,0)^{2} (3,2) (2,4)^{2} (5,1)^{2} (4,3)  $  \\ 
 & $(3,5) \ (6,2)^{2} \ (5,4) \ (0,8) \ (8,1) \ (4,6) \ (7,3) $ \\
 & $(8,4)$ \\
 &   \\
1 & $(0,1)^{3} \ (2,0)^{5} \ (1,2)^{12} \ (3,1)^{16} \ (0,4)^{7} \ (2,3)^{21} \ (5,0)^{12}$  \\
 & $(4,2)^{24} \ (1,5)^{13} \ (3,4)^{22} \ (6,1)^{20} \ (0,7)^{4} \ (5,3)^{22} \ (2,6)^{10}$ \\
  & $(8,0)^{8} \ (4,5)^{13} \ (7,2)^{16} \ (1,8)^{3} \ (6,4)^{11} \ (3,7)^{5} \ (9,1)^{7}$\\
  & $(5,6)^{6} \ (8,3)^{7} \ (2,9) \ (7,5)^{4} \ (11,0) \ (10,2)^{3} \ (6,7)$\\
  & $(9,4)^{2} \ (11,3)$\\
 &   \\
2 & $(0,0)^{9} \ (1,1)^{32} \ (3,0)^{27} \ (0,3)^{27} \ (2,2)^{93} \ (4,1)^{85} \ (1,4)^{70}$ \\ 
 & $(3,3)^{119} \ (6,0)^{62} \ (0,6)^{39} \ (2,5)^{80} \ (5,2)^{109} \ (4,4)^{115} \ (7,1)^{78}   $ \\
 & $(1,7)^{34} \ (3,6)^{55} \ (6,3)^{92} \ (5,5)^{61} \ (9,0)^{25} \ (0,9)^{4} \ (2,8)^{28}   $  \\
 & $(8,2)^{66} \ (7,4)^{44} \ (4,7)^{23} \ (10,1)^{25} \ (1,10)^{2} \ (6,6)^{25} \ (3,9)^{4}  $  \\
 & $(9,3)^{28} \ (5,8)^{3} \ (8,5)^{14} \ (12,0)^{7} \ (0,12) \ (11,2)^{10} \ (7,7)^{2}   $ \\
 & $(4,10) \ (10,4)^{9} \ (9,6) \ (13,1)^{2} \ (12,3)^{3} \ (11,5) \ (14,2)  $ \\
 &   \\
\hline
\end{tabular}
\caption{Space of the 6-$\alpha$-particle system, 
with the same observations as in Table \ref{tab2}.
}
\vspace{0.2cm}
\label{tab6alpha}
\end{table}
\end{center}

\begin{center}
\begin{table}[h!]
\centering
\begin{tabular}{|c|c|}
\hline
$n \hbar \omega$ & $7-\alpha$ \\
\hline
0 & $(0,0)^{2} \ (2,2)^{3} \ (4,1) (1,4) (3,3)^{3} (6,0)^{3} (0,6)^{3}  $  \\ 
 & $(2,5) \ (5,2) \ (4,4)^{4} \ (7,1) \ (1,7) \ (3,6)^{2} \ (6,3)^{2} $ \\
 & $(5,5)^{2} \ (8,2)^{2} \ (2,8)^{2} \ (6,6) / (9,3) \ (3,9) \ (12,0) $ \\
 & $(0,12)$ \\
 &   \\
1 & $(1,0)^{6} \ (0,2)^{8} \ (2,1)^{23} \ (1,3)^{29} \ (4,0)^{14} \ (3,2)^{44} \ (0,5)^{20}$  \\
 & $(2,4)^{43} \ (5,1)^{34} \ (4,3)^{53} \ (1,6)^{64} \ (7,0)^{70} \ (3,5)^{46} \ (6,2)^{36}$ \\
  & $(5,4)^{44} \ (0,8)^{10} \ (2,7)^{29} \ (8,1)^{20} \ (4,6)^{29} \ (7,3)^{27} \ (1,9)^{11}$\\
  & $(6,5)^{22} \ (3,8)^{15} \ (10,0)^{5} \ (9,2)^{15} \ (5,7)^{10} \ (8,4)^{12} \ (0,11)^{3}$\\
  & $(2,10)^{5} \ (7,6)^{6} \ (11,1)^{5} \ (4,9)^{4} \ (10,3)^{6} \ (6,8)^{2} \ (9,5)^{2} $\\
  & $(1,12) \ (3,11) \ (13,0) \ (8,7) \ (12,2)^{2} \ (5,10) \ (11,4) \ (14,1) $ \\
 &   \\
 2 & $(0,1)^{22} \ (2,0)^{68} \ (1,2)^{113} \ (3,1)^{163} \ (0,4)^{110} \ (2,3)^{225} \ (5,0)^{88}  $ \\
  & $(4,2)^{282} \ (1,5)^{189} \ (3,4)^{284} \ (6,1)^{185} \ (0,7)^{72} \ (5,3)^{284} \ (2,6)^{229}  $ \\
  & $(8,0)^{95} \ (4,5)^{249} \ (7,2)^{184} \ (1,8)^{100} \ (6,4)^{226} \ (3,7)^{159} \ (9,1)^{99} $ \\
  & $(5,6)^{146} \ (8,3)^{138} \ (0,10)^{33} \ (2,9)^{60} \ (7,5)^{115} \ (4,8)^{77} \ (11,0)^{23} $ \\
  & $(10,2)^{79} \ (6,7)^{54} \ (1,11)^{17} \ (9,4)^{64} \ (3,10)^{21} \ (8,6)^{38} \ (5,9)^{20} $ \\
  & $(12,1)^{26} \ (11,3)^{34} \ (7,8)^{11} \ (2,12)^{5} \ (10,5)^{16} \ (4,11)^{3} \ (14,0)^{8} $ \\
  & $(9,7)^{7} \ (6,10)^{4} \ (13,2)^{12} \ (12,4)^{9} \ (3,13) \ (8,9) \ (11,6)^{2} $ \\ 
  & $(15,1)^{5} \ (14,3)^{2} \ (10,8) \ (13,5) \ (17,0) \ (16,2) $ \\
 &   \\
\hline
\end{tabular}
\caption{Space of the 7-$\alpha$-particle system, 
with the same observations as in 
Table \ref{tab2}.
}
\vspace{0.2cm}
\label{tab7alpha}
\end{table}
\end{center} 

For the 3-$\alpha$-particle case (Table \ref{tab1}) the list 
is identical to the one published by Horiuchi \cite{horiuchi}.
In Table \ref{tab2} the space of the 4-$\alpha$-particle system
is compared to Kat${\bar {\rm o}}$'s list. 
In Table \ref{tab3}, the same
is done for the 5-$\alpha$-particle system
and in Tables \ref{tab6alpha} and \ref{tab7alpha} for the
6- and 7-particle case, respectively. 
While there
is no difference in the 3-$\alpha$ particle space within
the microscopic treatment and our approximation, the first 
difference appear in the 4-$\alpha$-particle case, from
$2\hbar\omega$ on. There, the irrep $(2,0)$ has a multiplicity 
of 2, while in \cite{kato} its is just one. As seen for
$3\hbar\omega$ the multiplicity in our approach is raising
significantly and also some additional irreps appear.
This increase of the multiplicity indicates that
many more states can be constructed, 
with the quantum numbers of an allowed shell model state
but not with the right symmetry of the $\alpha$-particles. 
However, these differences appear only for irreps with a
lower eigenvalue of the second Casimir operator and, thus,
are of lesser importance. One viable solution
is to restrict
to just multiplicity one, i.e., 
{\it implementing a further practical constraint},  
with the price to pay that
this eliminates additional states which are symmetric under
permutations. With this, the
problem of a too large multiplicity is avoided.
Note, that in 
a microscopic projection method the multiplicity also
increases with larger $n$ $\hbar\omega$ excitations. 

The 4-particle case is a good test for the quality of the
approximation, because exact procedures were
published in \cite{kramer} and the book \cite{mosh-smirnov}.
Unfortunately, this method is quite involved for
the 4-$\alpha$-particle system. In general, with the new 
method we get the same $SU(3)$-irreps, but with a larger
multiplicity with increasing
$n$ $\hbar\omega$. 

For the 5-$\alpha$-particle state 
the $0\hbar\omega$ list is identical
with the table of Kat${\bar {\rm o}}$, but for the $1\hbar\omega$ the 
discrepancies mentioned appear also for large $SU(3)$ irreps. 
In order to convince
the reader, in the Appendix we explicitly construct the
$0$ and $1\hbar\omega$ shell model space of $^{20}$Ne
and show that some states listed in \cite{kato} 
are indeed not part of the shell model space,
suggesting possible difficulties of the numerical procedure
applied in \cite{kato} in calculating the overlap
matrix elements (kernels). Eliminating those states,
a good agreement is reached again.

The present section also provides the new contribution
of 6- and 7-$\alpha$ particle states
up to large n $\hbar\omega$. The result is
useful for any cluster model which relies on a microscopic
Hilbert space, i.e., independent on what will
be presented in the next section. 
The procedure proposed can be directly
extended to a higher number of $\alpha$-cluster states. 

\section{Applications}
\label{appl}

The usefulness of the deduced approximate Hilbert space of 
$n$ $\alpha$-particle systems is investigated 
in this section, at hand of the SACM.
Results for $^{12}$C and $^{16}$O are resumed shortly, 
because these are already published in \cite{12C,16O}.
Also, the space for $^{12}$C is identical to the 
exact approach and for $^{16}$O differences only appear
from 2$\hbar\omega$ excitations on and there only for small
$SU(3)$ irreps, which do not have any sensible influence
at low energy.
Therefore, we restrict to systems with 5 ($^{20}Ne$),
6 ($^{24}$Mg) and 7 ($^{28}$Si) $\alpha$-particles.
In addition, instead of a full microscopic model,
the more easy to apply SACM 
\cite{cseh-letter,cseh-levai-anph} is chosen. The
applicability was investigated many times, were
we mention only one particular to a series of light nuclear
two-cluster systems \cite{sacm-appl1}. More than two
clusters are mostly not considered, which makes this
contribution also interesting from a conceptual point
of view.

There are, however, restrictions to the SACM making it 
difficult to treat certain aspects of a cluster system.
This is mainly related of not constructing a spatial
representation of the system. Also the states are stable
and cannot decay in a dynamical way, thus, thresholds
of cluster decays are difficult to simulate. Interesting
situations, like the transition to an $\alpha$-gas state,
are very difficult to study. Nevertheless, spectra, 
transition rates and spectroscopic factors can 
still be obtained.

\subsection{A particular SACM Hamiltonian,
electro-magnetic transition and 
spectroscopic factor operator}
\label{ham}

Here, not the most general Hamiltonian will be used but one
which satisfies the needs for light nuclei
\cite{12C,16O}, namely

\beqa
{\bd H} & = & \hbar\omega {\bd n}_\pi
+\chi (1-\chi_{n_\pi} \Delta {\bf n}_\pi )  {\bd {\cal C}}_2(\lambda , \mu ) 
+ t_3({\bd {\cal C}}_2(\lambda , \mu ))^2 
+ t_1 {\bd {\cal C}}_3(\lambda , \mu ) 
\nonumber \\
&& + \left(\xi+ \xi_{Lnp} \Delta {\bd n}_\pi \right) {\bd L}^2 +t_2 {\bd K}^2
\nonumber \\
&& +b_1 \left[\left({\bd\sigma}^\dagger \right)^2 
- \left( {\bd \pi}^\dagger \cdot {\bd \pi}^\dagger \right) \right]
\cdot \left[h.c.\right]
~~~.
\label{eq-6}
\eeqa
The first term defines 
the scale of the harmonic oscillator shell,
i.e., $\hbar\omega = 45~A^{-\frac{1}{3}} - 25~A^{-\frac{2}{3}}$
and it is fixed \cite{hw}, $A$ is the number of nucleons.
The ${\bd {\cal C}}_2(\lambda , \mu )$ is the 
second order Casimir
operator of $SU(3)$, which is proportional to the 
quadrupole-quadrupole interaction \cite{castanos}. The
${\bd K}^2$ term gives the square of the $K$-projection of the 
angular momentum onto the intrinsic z-axis 
and serves to distinguish states with the same
angular momentum within a given $SU(3)$ 
irrep \cite{pseudo-sympl}. The
$b_1$-term is proportional to a Casimir operator of $SO(4)$
and mixes $SU(3)$ irreps. The ${\bd L}^2$ is the angular 
momentum operator
with a factor simulating a variable moment of inertia.
The remaining terms are corrections. 

The $\widetilde{{\bd C}}_2\left({\widetilde \lambda},{\widetilde \mu}\right)$
is the second order Casimir-invariant of the coupled
$\widetilde{SU}(3)$ group, having contributions both from the internal
cluster part and from the relative motion. 
The $\widetilde{{\bd C}}_2\left({\widetilde \lambda},{\widetilde \mu}\right)$ is given by:

\begin{eqnarray}
\mbox{\boldmath $\widetilde{C}$}_2(\widetilde{\lambda},\widetilde{\mu}) & = & 2 \mbox{\boldmath $\widetilde{Q}$}^2 +
\frac{3}{4} \mbox{\boldmath $\widetilde{L}$}^2 ,  \nonumber \\
& \rightarrow & \left(\widetilde{\lambda}^2 + \widetilde{\lambda}\widetilde{\mu} 
+ \widetilde{\mu}^2 + 3\widetilde{\lambda} + 3\widetilde{\mu}
\right) ,  \nonumber \\
\mbox{\boldmath $\widetilde{Q}$} & = & \mbox{\boldmath $\widetilde{Q}$}_C 
+ \mbox{\boldmath $\widetilde{Q}$}_R ,
\nonumber \\
\mbox{\boldmath $\widetilde{L}$} & = & \mbox{\boldmath $\widetilde{L}$}_C 
+ \mbox{\boldmath $\widetilde{L}$}_R ,
\label{su3}
\end{eqnarray}
where $\mbox{\boldmath $\widetilde{Q}$}$ and 
$\mbox{\boldmath $\widetilde{L}$}$ are
the total quadrupole and angular momentum operators,
respectively, and $R$ refers to the relative motion. 
Thus, the second order Casimir operator describes the
quadrupole-quadrupole interaction strength and determines
the deformation of the nucleus.
Within the $SU(3)$-basis,
the eigenvalue of $\mbox{\boldmath $\widetilde{C}$}_2(\widetilde{\lambda},\widetilde{\mu})$ is also indicated.
The relations of the quadrupole and angular momentum
operators to the $\widetilde{C}^{(1,1)}_{2m}$ generators of the $\widetilde{SU}(3)$ 
group, expressed in terms of $\widetilde{SU}(3)$-coupled $\pi$-boson creation and
annihilation operators, are \cite{escher}:

\begin{eqnarray}
\mbox{\boldmath $\widetilde{Q}$}_{k,2m} & = & \frac{1}{\sqrt{3}}
\widetilde{C}^{(1,1)}_{k2m} , \nonumber \\
\mbox{\boldmath $\widetilde{L}$}_{k1m} & = & \widetilde{C}^{(1,1)}_{k1m} , 
\nonumber \\
\mbox{\boldmath $\widetilde{C}$}^{(1,1)}_{lm} & = & \sqrt{2} \left[
\mbox{\boldmath
$\pi$}^\dagger \otimes \mbox{\boldmath $\pi$} \right]^{(1,1)}_{lm} .
\label{su3gen}
\end{eqnarray}

The term with the square of the second order Casimir operator
serves to fine tune the relative positions of the band heads,
which is also the case for the third order Casimir operator,
whose eigenvalue within the $SU(3)$-basis is given by

\beqa
\widetilde{{\bf C}}_3 & \rightarrow & 
\left( \lambda - \mu \right) \left(2\lambda + \mu + 2 \right)
\left(\lambda + 2\mu + 3 \right)
~~~,
\label{c3}
\eeqa
These two correction terms improve slightly the adjustment of the band heads and are, therefore, of lesser importance.

The eigenvalue of the angular momentum operator
is $L(L+1)$ and of ${\bf K}^2$ is just $K^2$, where $K$
is the projection of the angular momentum on the 
intrinsic $z$-axis.

The $\chi_{n_\pi}$ term varies the
quadrupole-quadrupole interaction with increasing shell excitations
and the $\xi_{Lnp}$ term does the same for the moment of inertia,
which is inverse proportional to the factor of 
${\bd L}^2$.

The quadrupole electromagnetic transition operator is defined
as

\beqa
\bd{T}_m^{(E2)} & = & \sum_\gamma e_\gamma^{(2)}
\bd{Q}^{(2)}_{\gamma ,m}
~~~,
\label{trans-p}
\eeqa
where $e_\gamma^{(2)}$ is the effective charge of the
contribution to the quadrupole operator, coming from
the cluster $\gamma$ = $C_1$, $C_2$ and from the relative
motion $\gamma$ = $R$. A geometric estimate of
the effective charges is given in \cite{sacm-appl1}, thus,
the only parameter in the transition rates in an overall
factor, labeled $q_{\rm eff}$.

To resume: There are three parameters 
($\chi_{n_\pi}$, $t_1$, $t_3$)
which are responsible for the position of the band heads,
one parameter ($\chi_{n_\pi}$), which determines the
$\Delta n_\pi$ dependence of the quadrupole-quadrupole
interaction, two parameters which modify the
moment of inertia and $t_2$, which 
is responsible for the $K$-splitting. 
These parameters are adjusted {\it in average} 
to 13-18 energy
values, depending on the nucleus, from experiment.
The $B(E2)$ 
transition operator adds the overall scale parameter
$q_{eff}$. This scale factor is adjusted {\it in average} to 
2 $B(E2)$-values of the ground state band.

In \cite{specfac-draayer} a simple but effective 
algebraic expression for the spectroscopic factor 
for a {\it two cluster system} was proposed.
The spectroscopic factor is defined as

\beqa
&
S = 
& 
\nonumber \\
&
e^{{\cal A} + B n_\pi + C{\cal C}_2(\lambda_1,\mu_1) 
+ D{\cal C}_2(\lambda_2,\mu_2) 
+ E{\cal C}_2(\lambda_c,\mu_c)}
&
\nonumber \\
&
\times
e^{F{\cal C}_2(\lambda , \mu ) + G{\cal C}_3(\lambda , \mu ) 
+ H\Delta n_\pi}
&
\nonumber \\
&
\mid 
\langle (\lambda_1,\mu_1)\kappa_1L_1, 
(\lambda_2,\mu_2)\kappa_2L_2 \mid\mid
(\lambda_C,\mu_C)\kappa_C L_C \rangle_{\varrho_C}
&
\nonumber \\
& \cdot
\langle (\lambda_C,\mu_C)\kappa_CL_C, 
(n_\pi ,0)1l \mid\mid
(\lambda ,\mu )\kappa L \rangle_1
\mid^2
~~~,
&
\label{specfac-light}
\eeqa
where the $\varrho_C$-numbers refer to a 
multiplicity in the coupling to $SU(3)$ irreps and the $\kappa$'s to the 
multiplicities of the reduction of $SU(3)$ to $SO(3)$.
The parameters were adjusted to theoretically exactly calculated spectroscopic factors within the p- and sd-shell,
using the $SU(3)$ shell model \cite{draayer2}, with excellent results. For the good agreement, the factor depending on 
the $SU(3)$-isoscalar factors turns out to be crucial.
The parameters ${\cal A}$ to $H$, appearing in 
(\ref{specfac-light}) are listed in Table \ref{specpara}.

\begin{table}[]
	\begin{center}
		\begin{tabular}{|c|c|c|c|}
		\hline
${\cal A}$ & $B$ & $C$ & $D$  \\ 
\hline
- & -0.36113 & -0.054389 & -0.11764 \\
\hline
\hline
$E$ & $F$ & $G$ & $H$  \\ 
\hline
0.060728 & -0.0086654 & 0.000021097 & 1.9090 \\
\hline
\end{tabular}
	\caption{Parameter values used for the spectroscopic 
	factor of (\ref{specfac-light}).
	\label{specpara}}
	\end{center}
\end{table}

Because the expression in (\ref{specfac-light}) is only
valid for a two-cluster system, the results will correspond
for $^{20}$Ne to $^{16}$O+$\alpha$, for $^{24}$Mg to
$^{20}$Ne+$\alpha$ and for $^{28}$Si to 
$^{24}$Mg+$\alpha$.

\begin{table}[]
	\begin{center}
		\begin{tabular}{|c|c|c|c|}
		\hline
\textbf{Parameter} & $^{20}$Ne & $^{24}$Mg & $^{28}$Si  \\ 
\hline
$\hbar\omega$ & $13.19$ & $12.60$ & $12.11$  \\ \hline
$\chi$ & $-0.406628$ & $-0.177055$ & $-0.0737703$ \\ \hline
$\xi$ & $0.124941$ & $0.194327$ & $0.223200$  \\ 
\hline 
$t_1$ &  $0.00134669$ & $6.88122 \times 10^{-4}$ & $-3.45015 \times
10^{-3}$  \\ \hline 
$t_2$ & $-0.0333108$ & $0.314263$  & $-0.317268$  \\ 
\hline  		
$\chi_{n_\pi}$ & $-0.197490$ & $-0.0942936$ & $-0.0674217$ \\ 
\hline  		
$\xi_{L_{n_p}}$ & $-0.0473878$ & $-0.145655$ & $-0.16083$ \\
\hline
$b_1$ & $-0.350871$ & $0.188599$ & $-0.101639$ \\ 
\hline  		
$t_3$ & $0.00234264$ & $-6.31156\times 10^{-5}$ & $3.91536\times 10^{-4}$ \\
\hline
$q_{{\rm eff}}$ & 0.2930 & 0.290611 & 0.27301 \\ 
\hline  		
\end{tabular}
	\caption{Non-zero parameter values for 
	$^{20}$Ne, $^{24}$Mg and $^{28}$Si.
	\label{parameters}}
	\end{center}
\end{table}

\begin{table}[]
	\begin{center}
		\begin{tabular}{|c|c|c|c|c|c|c|}
		\hline
$J_i^{P_i} \rightarrow J_f^{P_f}$ & $^{20}$Ne Th & 
$^{24}$Mg th & $^{28}$Si th & $B(E\_2)[Ne]$ & $B(E\_2)[Mg]$ & $B(E\_2)[Si]$ \\ \hline  
$2^+_1 \rightarrow 0^+_1$ & $21.78$ & $25.97$ & $15.54$ & 
$20.3 \pm 1$ & $21.5 \pm 1.0$ & $13.2 \pm 0.5$ \\ \hline 
$2^+_1 \rightarrow 0^+_2$ & $0.38$ & $0.095$  & $0$  & $0.027$ & $0.128 \pm 0.02$ & - \\ \hline
$2^+_1 \rightarrow 0^+_3$ & $0$ & $0.0013$ & $0$ & $-$ & 
$-$ & $0.053 \pm 0.003$ \\ \hline 
$2^+_2 \rightarrow 0^+_1$ & $37.79$ & $2.29$ & $0.012$ & $-$ & 
$1.94 \pm 0.19$ & $0.37 \pm 0.15$  \\ \hline 
$2^+_2 \rightarrow 0^+_2$ & $72.09$ & $0.071$ & $5.92$ & - & $1.78 \pm 0.28$ & $0.8 \pm 0.5$ \\ \hline  
$2^+_2 \rightarrow 0^+_3$ & $0$ & $7.87 \times 10^{-4}$ & $0$ & - & $-$ & $-$ \\ \hline  
$2^+_2 \rightarrow 2^+_1$ & $0$ & $8.96$ & $2.6 \times 10^{-5}$ & $0.73 \pm 0.09$ & $-$ & $-$ \\ \hline  
$4^+_1 \rightarrow 2^+_1$ & $20.78$ & $36.13$ & $20.55$ & $22 \pm 2$ &  $39 \pm 4$ & $16.4 \pm  1.8$ \\ \hline 
$2^+_3 \rightarrow 0^+_1$ & $0$ & $0.0035$  & $0.023$ & $0.73$ & $0.67 \pm 0.23$ & $0.162 \pm 0.018$ \\ \hline 
$2^+_3 \rightarrow 0^+_2$ & $0$ & $16.90$  & $7.56$  & $-$ & $0.36 \pm 0.004$ & - \\ \hline
$4^-_1 \rightarrow 2^-_1$ & $0$ & $57.73$ & $47.15$ & $1.8$ & - & - \\ \hline 
$4^-_1 \rightarrow 3^-_1$ & $0$ & $27.54$ & $18.41$ & - & - & $0.91 \pm 0.19$ \\ \hline 
		\end{tabular}
	\caption{
	Theoretical $B(E2)$-transition values 
	of the systems $^{20}$Ne,
	$^{24}$Mg and
	$^{28}$Si. The unit is in WU and the
	theoretical values are
	compared to available experimental data \cite{brookhaven}.
	When the theoretical values is $10^{-7}$ or less, the 
	tabulated value is set to zero.}
	\label{BE2}
	\end{center}
\end{table}

\begin{center}
\begin{table}[h!]
\centering
\begin{tabular}{|c|c|c|c|}
\hline\hline
%ind&&
$J_k^P$ & $^{20}$Ne (th) & $^{24}$Mg (th)
& $^{28}$Si \\
\hline
$0_1^+$ & 0.2248 & 0.0001928 &  $1. \times 10^{-6}$ \\
$0_2^+$ & 0.1027 & 0.0002546 &  $2 \times 10^{-6}$ \\
$2_1^+$ & 0.2241 & 0.00007083 & $2 \times 10^{-7}$ \\
$2_2^+$ & 0.1034 & 0.00001039 & $3 \times 10^{-7}$ \\
$4_1^+$ & 0.2217 & $2.325 \times 10^{-6}$ & $7 \times 10^{-8}$ \\
$4_2^+$ & 0.1057 & 0.00008475 & $4 \times 10^{-7}$ \\
$1_1^-$ & 0.1365 & 0.00001498 & $8 \times 10^{-8}$ \\
$3_1^-$ & 0.1365 & 0.00005610 & $3 \times 10^{-7}$ \\
$5_1^-$ & 0.1365 & 0.00005075 & $1 \times 10^{-7}$ \\
\hline
\hline 
 \end{tabular}
\caption{
Some spectroscopic factors, divided by $e^{\widetilde{A}}$, of low lying states. 
In the first column the state considered is listed.
The spectroscopic factors correspond to a two cluster
system of the type $X+\alpha$, where $X$ is a also
a multi-$\alpha$-particle state.
} 
\vspace{0.2cm}
\label{Spec}
\end{table}
\end{center}

In order to compare some data available in $^{20}$Ne,
an alternative definition of the spectroscopic factor
is used, namely the 
{\it dimensional reduced $\alpha$-width} $\theta_\alpha^2$
\cite{cseh1991}

\beqa
\theta_\alpha^2 & = & \frac{\gamma_\alpha^2}{\gamma_W^2}
~~~,
\label{reduced}
\eeqa
where $\gamma_\alpha^2$ is the reduced width and
$\gamma_W^2$ is the {\it Wigner limit}

\beqa
\gamma_W^2 & = & \frac{3\hbar^2}{2\mu a^2}
~~~,
\label{wigner}
\eeqa
with $\mu$ as the reduced mass of the two cluster system
and $a$ the channel radius within the R-matrix formulation.

\subsection{$^{12}$C and $^{16}$O summary}
\label{12c-16o}

In \cite{12C,16O} the $^{12}$C nucleus, as a three
$\alpha$-particle state, and $^{16}$O, as a four
$\alpha$-particle state, where investigated within the
SACM, with the motivation to study the role of the
{\it Pauli Exclusion Principle} (PEP). The space for
$^{12}$C was copied from \cite{horiuchi} and the space
for $^{16}$O was copied from \cite{kato}. These spaces
agree with the ones constructed with the method 
proposed in this contribution, except for the 
multiplicity of one state at $2\hbar\omega$, in the
$4\alpha$ particle system and also in higher shells.

The main point is that the PEP is very important in these
systems, leading otherwise to a wrong structure
of states at low energy and erroneous interpretations,
even when the main degrees of freedoms are taken correctly
into account. This shows that having identifies the correct
number of degrees of freedom is not sufficient and that
the correct Hilbert space is as important, if
not even more.

Observing the PEP, using the SACM leads to a satisfying
interpretation of the spectrum so far measured. The SACM
shows that other models which take not into account
the PEP will lead to a too dense spectrum at low energy.
Future experimental studies surely will confirm this,
simply because the PEP is a real and important principle
of nature.

The $^{12}$C and $^{16}$O nuclei were investigated in relation
to the structure of Hoyle states \cite{schuck2018}, as the
possible formation of $\alpha$-gas 
like states \cite{schuck2013}. The main theory is
described in \cite{japan2017}, which uses ab initio 
methods.
In \cite{fukui} the spatial manifestation of $\alpha$
condensation was investigated, related to a phase
transition to an $\alpha$-gas condensate. 
Here, we have to emphasize that the SACM is not suited
well for the description of an $\alpha$ gas, also 
due to the present simplified structure of the Hamiltonian.

\subsection{$^{20}$Ne}
\label{20Ne}

One of the rare applications of cluster models
to $^{20}$Ne is published in \cite{adachi2020}, 
were an experimental candidates of a 5-$\alpha$ particles 
was reported and 
the interpretation relies on \cite{japan2017}.
These multi-$\alpha$-particle states were related to 
high lying $0^+$ states. Interesting enough is that the
position and sequence of the $0^+$ states show a similar
behavior in our calculations as in \cite{japan2017}:
While the $0^+$ states in \cite{japan2017} follow the
sequence 
$E(0_2^+)=6.06$MeV, $E(0_3^+)=11.26$MeV,
$E(0_4^+)=12.05$MeV, $E(0_5^+)=14.03$MeV, 
$E(0_6^+)=14.03$MeV, 
in our compilation the sequence is
$E(0_2^+)=6.69$MeV, $E(0_3^+)=8.61$MeV,
$E(0_4^+)=11.95$MeV, $E(0_5^+)=12.01$MeV, 
$E(0_6^+)=12.92$MeV, i.e., the same states of
$0_5^+$ and $0_6^+$ are almost degenerate.

Unfortunately, the model presented
here, is too simple for giving realistic contributions
to the structure of and $\alpha$ gas. However, the 
multi-$\alpha$-particle states constructed in Section 
\ref{general},
serve as a practical basis when an ab initio model is used.
In \cite{rickards1984} some reduced $\alpha$ widths were
obtained. Within the SACM these values cannot
be compared directly, due to the missing scale factor or
the nearly $SU(3)$ symmetry results in zero value.
For example, the reduced $\alpha$ width for the 
$0_2^+$ state is $0.17 \pm 0.08$, while we obtain
$0.10 e^{\cal A}$, where $e^{\cal A}$ is the scaling factor.
From there, adjusting to the deduced value we obtain
$e^{\cal A}=1.7$. For the $2_2^+$ and $4_2^+$ state the
experimentally deduced values are 0.047 and 0.17, 
respectively, while we obtain the same value as for 
$0_1^+$, resulting in the prediction of $0.17$, quite
well reproducing the experimental observation. 

\begin{figure}
\centerline{
\rotatebox{0}{\resizebox{400pt}{400pt}{\includegraphics[width=0.23\textwidth]{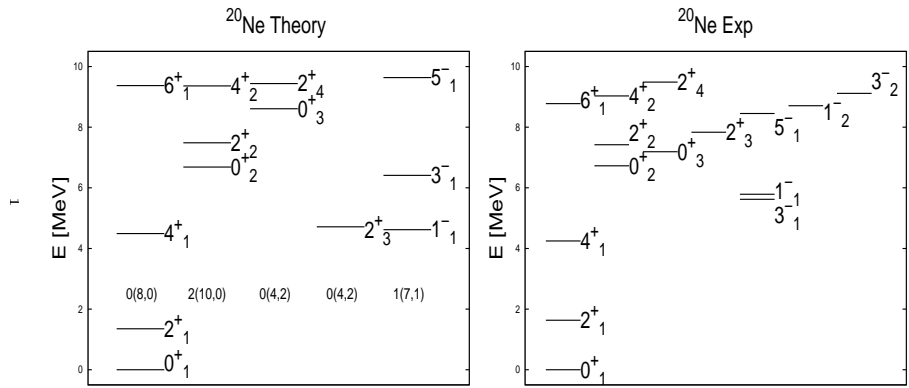}}}
%\rotatebox{270}{\resizebox{150pt}{200pt}{\includegraphics[width=0.23\textwidth]{20Ne_the.pdf}}}
%\rotatebox{270}{\resizebox{150pt}{200pt}{\includegraphics[width=0.23\textwidth]{20Ne_exp.pdf}}}
}
\caption{\label{20-Ne} 
Spectrum of $^{20}$Ne, as a 5-$\alpha$-particle states.
The theoretical spectrum
(left panel) is compared to experiment (right panel).
The theoretical spectrum was obtained
including $SO(4)$ mixing. 
The dominant $\Delta n_\pi(\lambda , \mu )$ 
quantum numbers are indicated
below each band, where $\Delta n_\pi$ refers
to the number of shell excitations. 
Experimental data are retrieved from \cite{brookhaven}.
}
\end{figure}

The model Hamiltonian presented above is used to fit the spectrum. Doing so, we have to take into account that
$^{20}$Ne is a special nucleus, because only those states are
considered which are reached in the 
$^{16}$O$\left(\alpha , \gamma \right)$
reaction channel. Especially some low-lying negative 
parity states are not observed in the above mentioned
reaction, which makes it difficult to associate the remaining
states into bands and compare to the theory.

In Fig. \ref{20-Ne} the adjusted spectrum of $^{20}$Ne is depicted. On the left hand side, 
the spectrum is depicted, as obtained with a small $SO(4)$
mixing (i.e., it is essentially $SU(3)$).
The position of the band heads is well reproduced.
Considering the complex structure of the spectrum and
the simple model Hamiltonian, the agreement is satisfactory.
In Table \ref{BE2} some $B(E2)$-values are listed.

Finally, in Table \ref{Spec} some spectroscopic factors
of $^{16}$O+$\alpha$ $\rightarrow$ $^{20}$Ne are tabulated.
Also here we see, that $^{20}$Ne is special, because 
several spectroscopic
factors different from zero are found. We did also the calculation
in the $SU(3)$-limit, where the parameter $b_1$ mixing with $SO(4)$
is set to zero. The values of the spectroscopic factors different 
from zero did not change, nor other non-zero values appeared.
The main reason is that the cluster irrep is $(0,0)$ (the
$^{16}$O cluster and the $\alpha$ cluster carry both the
scalar irrep) the relative irrep is $(8,0)$, such that for
$0\hbar\omega$ only the $(8,0)$ irrep can be reached
and for $2\hbar\omega$ only the $(10,0)$ irrep can
be reached. Due to this characteristic
some irreps cannot be reached, as for example the
$(4,2)$ irrep at $0\hbar\omega$ has according to Eq. 
\ref{specfac-light} a zero spectroscopic factor.

\subsection{$^{24}$Mg}
\label{24Mg}

This nucleus is not often considered in 
the literature, except for example in 
\cite{schuck2013}, where even the concept of a
$n$ $\alpha$-particle state was discussed, for up
to $n=10$. Experimental evidence of a 7-$\alpha$-particle,
related to excited states in $^{56}$Fe, was reported in
\cite{akimune2013}. In \cite{descou2021} the cluster
$\alpha$-structure of $^{24}$Mg 
(besides the one of $^{12}$C)
is investigated using a microscopic cluster approach,
with a satisfying agreement. However, 
instead of a 6-$\alpha$-particle system the
$^{18}$O+$\alpha$+$\alpha$ was considered. 
In this model, the Hoyle
equivalent states show a significant larger 
root-mean-square radius. In $^{24}$Mg two
$0^+$ states are identified just under the three-body 
threshold, corresponding to 
Hoyle equivalent states.

Also for $^{24}$Mg the model Hamiltonian of the SACM
is too simple to describe these states, but the
microscopic model space constructed in this contribution
serves as a basis for the description of these states
within a full microscopic model.

\begin{figure}
\centerline{
\rotatebox{0}{\resizebox{400pt}{400pt}{\includegraphics[width=0.23\textwidth]{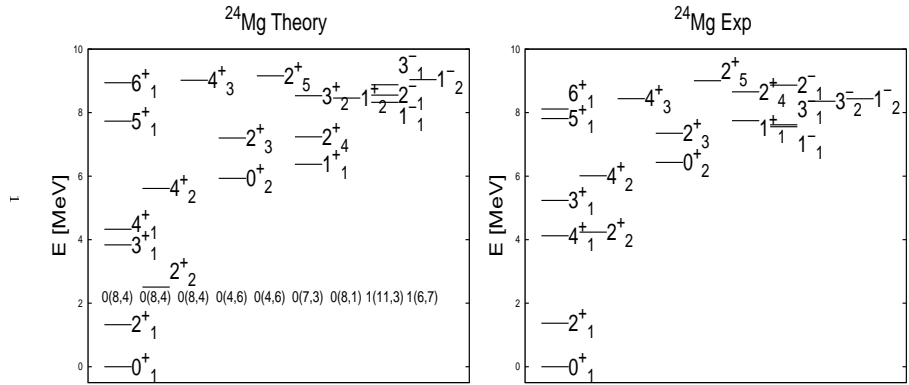}}}
%\rotatebox{270}{\resizebox{150pt}{200pt}{\includegraphics[width=0.23\textwidth]{24Mg_the.pdf}}}
%\rotatebox{270}{\resizebox{150pt}{200pt}{\includegraphics[width=0.23\textwidth]{24Mg_exp.pdf}}}
}
\caption{\label{24-Mg} 
Spectrum of $^{24}$Mg, as a 6-$\alpha$-particle states.
The theoretical spectrum
(left panel) is compared to experiment (right panel).
The theoretical spectrum was obtained,
including $SO(4)$ mixing.
The dominant $\Delta n_\pi(\lambda , \mu )$ quantum number 
are indicated below each band, where $\Delta n_\pi$
refers to the number of shell excitations. 
The dominant $\Delta n_\pi(\lambda , \mu )$ 
quantum numbers are indicated
below each band, where $\Delta n_\pi$ refers
to the number of shell excitations. 
Experimental data are retrieved from \cite{brookhaven}.
}
\end{figure}

For the $^{24}$Mg nucleus all reaction channels are taken into
account and the experimental energies are retrieved from
\cite{brookhaven}.
In Fig. \ref{24-Mg} the adjusted spectrum, compared to the
experiment, is depicted. 
On the left hand side, 
the spectrum, as obtained with $SO(4)$
mixing, is depicted.
In Table \ref{BE2} the $B(E2)$ values are listed and
also here the agreement to experiment is quite well.

Finally, In Table \ref{Spec} predictions for some 
spectroscopic factors
of $^{20}$Ne+$\alpha$ $\rightarrow$ $^{24}$Mg are tabulated.

\subsection{$^{28}$Si}
\label{28Si}

For $^{28}$Si we did not find a relevant detailed discussion,
except to the mention of a multi-$\alpha$ particle
state in \cite{schuck2013}. Nevertheless, the results
presented here and the microscopic model space
constructed, may be of interest.

\begin{figure}
\centerline{
\rotatebox{0}{\resizebox{400pt}{400pt}{\includegraphics[width=0.23\textwidth]{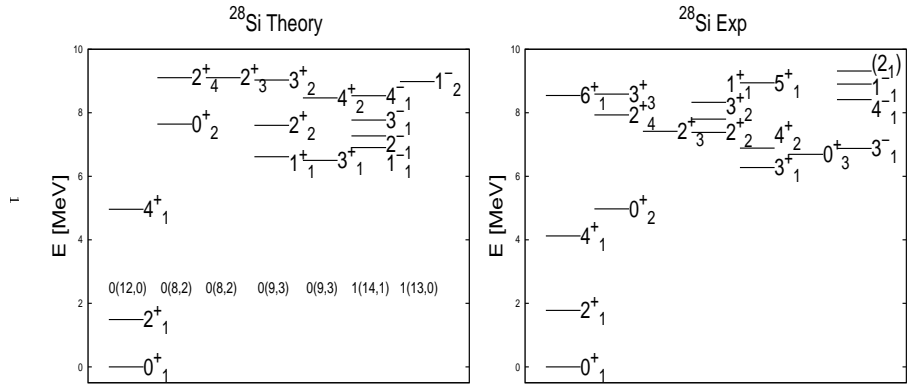}}}
%\rotatebox{270}{\resizebox{150pt}{200pt}{\includegraphics[width=0.23\textwidth]{28Si_the.pdf}}}
%\rotatebox{270}{\resizebox{150pt}{200pt}{\includegraphics[width=0.23\textwidth]{28Si_exp.pdf}}}
}

\caption{\label{28-Si} 
Spectrum of $^{28}$Si, as a 7-$\alpha$-particle states.
The theoretical spectrum
(left panel) is compared to experiment (right panel).
On the left hand side the theoretical spectra, obtained
including $SO(4)$ mixing, is depicted, compared to experiment (right hand side). 
The dominant $\Delta n_\pi(\lambda , \mu )$ quantum numbers 
are indicated below each band, where $\Delta n_\pi$
refers to the number of shell excitations. 
Experimental data are retrieved from \cite{brookhaven}.
}
\end{figure}

This nucleus is in the middle of $sd$-shell and is already
at the limit of validity of the $SU(3)$ model, due to the
increase of the spin-orbit interaction. Nevertheless,
it may still serve as a good example of a multi-$\alpha$ 
particle system and simultaneously an approximate
$SU(3)$ symmetry.

In Fig. \ref{28Si} the calculated spectrum is shown, with
a small
$SO(4)$ mixing, on the right hand side, compared to the
experimental spectrum on the left hand side. 
The parameters, used in the $SO(4)$-mixing,
are listed in Table \ref{parameters}.
Some calculated $B(E2)$-values, compared to experiment,
are listed in Table \ref{BE2} and the spectroscopic 
factors in the $SU(3)$ approximation are listed in
Table \ref{Spec}. Note that the spectroscopic factors have
decreased even more with respect to the other systems.
In general, the heavier the nucleus becomes, the stronger
the value of the spectroscopic factor decreases,
Though, the $^{28}$Si
nucleus is a limiting case, the SACM still can describe it
fairly well.

\section{Conclusions}
\label{concl}

The main emphasis of this contribution is on the approximate
method proposed for the construction of the shell model space for 
an arbitrary multi-$\alpha$-particle system. Though, the
resulting states are allowed by the {\it Pauli Exclusion principle}, not all are symmetric under interchange of
the $\alpha$-particles, which is due to the approximation
applied. The method is practical and easy to perform for
any number of the $\alpha$-particles.
This is one reason to accept the errors, which appear at
higher shell excitations and also are related to
$SU(3)$ irreps with smaller eigenvalues of the second
order Casimir operator, i.e., higher lying states.
Apart from the 3-, 4- and 5-$\alpha$-particle spaces,
as a new result we obtained the Hilbert space
of 6- and 7-$\alpha$ particles.
The agreement to formerly calculated spaces are good,
with a disagreement to \cite{kato} for the 5-$\alpha$ particle case
and 1$\hbar\omega$ excitation. However, we could prove that
that the low lying 
states, listed in \cite{kato}, do not appear in the 
shell model space. 

We hope that the procedure exposed is of great help
not only for an algebraic model, but microscopic 
theories as well, where the construction of the model space, using numerical procedures, are more involved
and/or prohibitive.

The model space was applied within the {\it Semimicroscopic
Algebraic Cluster Model} (SACM) 
to $^{20}$Ne, $^{24}$Mg and
$^{28}$Si, after a short resum\'e for $^{12}$C
and $^{16}$O was presented. The obtained theoretical spectra
and transition values were in fair agreement to experiment.
The predicted spectroscopic factors for the $X+\alpha$ 
system, with
$X$ being a multi-cluster $\alpha$ state, show a steady
decline with increasing number of $\alpha$ particles.
These results may also serve as comparison for more
detailed and involved models of a multi-$\alpha$-particle
system \cite{schuck2018}.

\section*{Appendix: Construction of the 
$1\hbar\omega$ shell model space
for the 5-$\alpha$ cluster system}
\label{app}

In constructing the $5-\alpha$-particle system, for
the $SU(3)$ content
of Pauli allowed states
a discrepancy was found to \cite{kato} for the $1\hbar\omega$
excitation.
The states of the 
$1\hbar\omega$ excitation $(10,4)$, $(8,5)$, $(9,3)$, 
$(7,4)$ and
$(5,5)$, appearing in the Table II of \cite{kato},
{\it do not appear in the shell model space we constructed}.

Therefore, in what follows
we construct the complete shell model
space of $^{20}$Ne for $1\hbar\omega$. Because
Kat${\bar {\rm o}}$ \cite{kato} claims that the states constructed
are Pauli allowed, they should appear in this shell model
space.

First, the general procedure will be explained and
afterward the shell model space for $U_{ST}(4)$ irreps
mostly antisymmetric is constructed. Non-mostly 
antisymmetric states belong to higher lying supermultiplets.
Nevertheless, at the end all remaining irreps are 
constructed, i.e., the complete shell model space. 
We will show that the above mentioned irreps
are not contained in the complete $1\hbar\omega$
shell model space and, thus, cannot be anti-symmetric.
 
\subsection*{General consideration}
\label{general2}

The classification within each shell $\eta$ is given by

\beqa
U((\eta +1)(\eta +2)) & \supset 
U(\frac{1}{2}(\eta +1)(\eta +2))~ \otimes & U_{ST}(4)
\nonumber \\
\left[ 1^N \right] ~~~~~~& ~~~ \left[\tilde{h}\right] &
~~~\left[ h \right]
\nonumber \\
& ~~~ \cup &
\nonumber \\
& ~~~~~~~~~~~~~~~~~~ SU(3) ~\varrho(\lambda , \mu)
~~~, 
\label{eq-1}
\eeqa
where $N$ is the number of particles in this shell,
$\left[ h \right]$ = $ \left[ h_1h_2h_3h_4\right]$
is the irrep of the spin-isospin group $U_{ST}(4)$
and $\left[ \tilde{h}\right]$ is the conjugate
(interchanging rows with columns) of $\left[ h \right]$.
The $U(\frac{1}{2}(\eta +1)(\eta + 2))$ group is reduced 
to $SU(3)$, with $(\lambda , \mu )$ referring to the $SU(3)$
irrep with $\varrho$ its multiplicity. The routines
developed in \cite{bahri} are used.

One usually restricts to $U_{ST}(4)$ irreps of the form
$[hhhh]$, $[(h+1)hhh]$, $[(h+1)(h+1)hh]$ or 
$[(h+1)(h+1)(h+1)h]$, which are due to the supermultiplet
structure in nuclei. We can, though, consider also all other
possibilities, which will be done later, at the second part
of this Appendix.

\vskip 0.5cm
\noindent
{\bf Procedure:}

\begin{itemize}

\item Distribute the nucleons in the shell, starting with 
$0\hbar\omega$.

\item En each shell $\eta$, fix the $U_{ST}(4)$ irrep, which
determines the $U(\frac{1}{2}(\eta +1)(\eta +2))$ irrep
and reduce to $SU(3)$.

\item For a given $n\hbar\omega$ excitation, multiply the
results of all shells.

\item Suppose we got in this manner the content of the 
$n\hbar\omega$ excitation. 
Reduce the center of mass by multiplying the result
for $0\hbar\omega$ by $(n,0)$, the result of $1\hbar\omega$ with
$(n-1,0)$, etc., until reaching $(n-1)\hbar\omega$, whose
result has to be multiplied by $(1,0)$. The such obtained
list of irreps all correspond to
states containing contributions of the center
of mass motion and have to be subtracted from the list
of multiplying all $SU(3)$ irreps from different shells.

\item {\bf The so obtained list of $SU(3)$ irreps
correspond to the shell model space of $n\hbar\omega$.
The complete space is obtained by using all other
possible $U_{ST}(4)$ irreps and repeating the procedure above.
}

\item The list of Kat${\bar {\rm o}}$ \cite{kato} 
should have an overlap to the 
calculated shell model space. No irrep is allowed to only
appear in the list provided by Kat${\bar {\rm o}}$ but not 
in the 
shell model space!

\end{itemize}

\section*{A: Shell model space for the most ant-symmetric
irrep of $U_{ST}(4)$}

In this section the shell model space, using only the
most antisymmetric irrep of $U_{ST}(4)$ is constructed.
These $U_{ST}(4)$ irreps are the lowest lying states in
energy. In a subsequent section, all the remaining irreps
are added, for completeness.

\subsection*{$0\hbar\omega$ space}
\label{0hw}

Here, $\eta = \eta_v$, for $\eta_v$ the valence shell.
The s-shell can be neglected because it is full and 
contributes only with a $(0,0)$ irrep.

In case of $^{20}$Ne, the 5-$\alpha$-particle system,
there are 4 nucleons in the $\eta = 2$ (sd) shell.
The $U_{ST}(4)$ irrep in the sd-shell is $[1^4]$ and thus
$[4]$ for U(6). The $SU(3)$ content of this irreps is
(we use here and for all reductions the $SU(3)$
library, published in \cite{bahri}):

\beqa
(8,0)~+~(4,2)~+~(0,4)~+~(2,0)
~~~.
\label{eq-2}
\eeqa
This agrees with the table listed in Kat${\bar {\rm o}}$'s publication
in 1988 \cite{kato}.

\subsection*{$1\hbar\omega$ space}
\label{1hw}

There are two cases of distributing the nucleons 
in the shells:

\vskip 0.5cm
\noindent
a) 4 nucleons in the s-shell, 12 in the p-shell,
3 in the sd shell and 1 in the pf-shell. The s- and p-shell
do not contribute, because they are closed.\\
b) 4 nucleons in the s-shell, 11 (one hole) in the p-shell
and 5 nucleons in the sd-shell.

\vskip 0.5cm
{\it These are all possibilities for the $1\hbar\omega$
excitation.}

\subsubsection*{Case a)}
\label{case-a}

The $SU(3)$ content of 3 particles in the sd shell and 
of one in the pf-shell, respectively, is 
(the irrep of $U_{ST}(4)$ is given by 
$[1^3]_{U_{ST}(4)}$, which corresponds to $[3]$ in U(6)),

\beqa
{\rm sd}:&&~~~(6,0)~+~(2,2)~+~(0,0)
\nonumber \\
{\rm pf}:&&~~~(3,0)
~~~.
\label{eq-3}
\eeqa

Multiplying the result of the sd-shell with the pf-shell,
we obtain:

\beqa
(6,0) \otimes (3,0) & = & (9,0)~+~(7,1)~+~(5,2)~+~(3,3)
\nonumber \\
(2,2) \otimes (3,0) & = & (5,2)~+~(3,3)~+~(1,4)~+~
(4,1)
\nonumber \\
&& ~+~(2,2)~+~(0,3)~+~(3,0)~+~(1,1)
\nonumber \\
(0,0) \otimes (3,0) & = & (3,0)
\label{eq-4}
\eeqa

\subsubsection*{Case b)}
\label{case-b}

The $SU(3)$ content of 5 particles in the sd shell 
(we have the $U_{ST}(4)$ irrep $[2,1^3]$ and 
therefore in $\eta=2$ the orbital irrep $[4,1]_{U(6)}$) and 
of one hole in the p-shell, respectively, is

\beqa
{\rm sd}:&&~~~(8,1)~+~(6,2)~+~(4,3)~+~(5,1)~+~(2,4)~+~
(3,2)~+~(4,0)
\nonumber \\
&& ~+~(1,3)~+~(2,1)~+~(0,2)
\nonumber \\
{\rm p}:&&~~~(0,1)
~~~.
\label{eq-3a}
\eeqa

Multiplying the result of the sd-shell with the p-shell,
we obtain:

\beqa
(8,1) \otimes (0,1) & = & (8,2)~+~(9,0)~+~(7,1)
\nonumber \\
(6,2) \otimes (0,1) & = & (6,3)~+~(7,1)~+~(5,2)
\nonumber \\
(4,3) \otimes (0,1) & = & (4,4)~+~(5,2)~+~(3,3)
\nonumber \\
(5,1) \otimes (0,1) & = & (5,2)~+~(6,0)~+~(4,1)
\nonumber \\
(2,4) \otimes (0,1) & = & (2,5)~+~(3,3)~+~(1,4)
\nonumber \\
(3,2) \otimes (0,1) & = & (3,3)~+~(4,1)~+~(2,2)
\nonumber \\
(4,0) \otimes (0,1) & = & (4,1)~+~(3,0)
\nonumber \\
(1,3) \otimes (0,1) & = & (1,4)~+~(2,2)~+~(0,3)
\nonumber \\
(2,1) \otimes (0,1) & = & (2,2)~+~(3,0)~+~(1,1)
\nonumber \\
(0,2) \otimes (0,1) & = & (0,3)~+~(1,1)
~~~.
\label{eq-4a}
\eeqa

\subsubsection*{Intermediate result for the $1\hbar\omega$
excitation}
\label{inter}

Summing the result for case a) and b), we obtain

\beqa
&
(9,0)^2~+~(7,1)^3~+~(3,3)^5~+~(1,4)^3~+~(4,1)^4
&
\nonumber \\
&
~+~(2,2)^4~+~(0,3)^3~+~(3,0)^4~+~(1,1)^3~+~(8,2)~+~(6,3)
&
\nonumber \\
&
~+~(5,2)^5~+~(4,4)~+~(6,0)~+~(2,5)
~~~,
\label{eq-5}
\eeqa
where an upper index denotes the multiplicity.

\subsubsection*{Removing the center of mass}
\label{cm}

For this, we multiply all irreps in the $0\hbar\omega$
space, given in (\ref{eq-2}), by $(1,0)$:

\beqa
(8,0) \otimes (1,0) & = & (9,0)~+~(7,1)
\nonumber \\
(4,2) \otimes (1,0) & = & (5,2)~+~(3,3)~+~(4,1)
\nonumber \\
(0,4) \otimes (1,0) & = & (1,4)~+~(0,3)
\nonumber \\
(2,0) \otimes (1,0) & = & (3,0)~+~(1,1)
~~~.
\label{eq-6a}
\eeqa

All the irreps on the right hand side represent
irreps with center of mass contributions and have
to be subtracted from the list in (\ref{eq-5}).

\subsection*{Final result for the $1\hbar\omega$
excitation} 
\label{final}

Subtracting (\ref{eq-6a}) from (\ref{eq-5}) leads to
the final list

\beqa
&
(9,0)~+~(7,1)^2~+~(3,3)^4~+~(1,4)^2~+~(4,1)^3
&
\nonumber \\
&
~+~(2,2)^4~+~(0,3)^2~+~(3,0)^3~+~(1,1)^2~+~(8,2)~+~(6,3)
&
\nonumber \\
&
~+~(5,2)^4~+~(4,4)~+~(6,0)~+~(2,5)
~~~.
\label{eq-7}
\eeqa
This list has to be compared with the one of
Kat${\bar {\rm o}}$'s table \cite{kato} for 5 $\alpha$ particles. The list
{\it does not contain the irreps}

\beqa
(10,4)~,~(8,5)~,~(9,3)~,~(7,4)~,~(5,5)
~~~.
\label{eq-8}
\eeqa
Thus, for example $(10,4)$ does not have an overlap with
the shell model space and cannot satisfy the Pauli 
exclusion principle. All others are also not contained, 
restricting to the {\it most anti-symmetric irrep
of $U_{ST}(4)$ en each shell}. If they are contained
using other $U(4)$ irreps will be investigated
in the next section. But minimally
$(10,4)$ {\it is not contained}, because the other 
$U_{ST}(4)$ irreps will produce $SU(3)$ irreps which are
not as large as $(10,4)$.

\section*{B: Completing the shell model space}
\label{compl}

In this section we will include further irreps of
$U_{ST}(4)$, which correspond to supermultiplets at
higher energy.

\subsubsection*{$0\hbar\omega$ excitation}
\label{0hw-2}

The four missing $U_{ST}(4)$ irreps are 
$[2,1^2]_{U_{ST}(4)}$, $[3,1]_{U_{ST}(4)}$,
$[2^2]_{U_{ST}(4)}$ and
$[4]_{U_{ST}(4)}$, which imply a $U(6)$ representation (sd-shell)
of $[3,1]_{U(6)}$, $[2,1^2]_{U(6)}$, $[2^2]_{U(6)}$
and $[1^4]_{U(6)}$, respectively.

The reduction of the $U(6)$ irreps to $SU(3)$ is

\beqa
[3,1]_{U(6)} & \rightarrow & (6,1)~+~(4,2)~+~(2,3)
~+~(3,1)~+~(1,2)
\nonumber \\
&& ~+~ (2,0) 
\nonumber \\
\left[2,1^2\right]_{U(6)} & \rightarrow & 
(5,0)~+~(2,3)~+~(3,1)~+~(1,2)~+~(0,1)
\nonumber \\
\left[2^2\right]_{U(6)} & \rightarrow & (4,2)~+~(3,1)~+~(0,4)~+~(2,0)
\nonumber \\
\left[1^4\right]_{U(6)} & \rightarrow & (1,2)
~~~.
\label{eq-9}
\eeqa

\subsubsection*{$1\hbar\omega$ excitation}
\label{1hw-2}

We use the same notation for case a) and case b).

\subsubsection*{Case a)}
\label{case-1-2}

The $\left[3\right]_{U(6)}$ was already considered in the
former section. The remaining irreps of $U(6)$ are
$\left[2,1\right]_{U(6)}$ and $\left[1^3\right]_{U(6)}$

The reduction of the $U(6)$ irrep is:

\beqa
{\rm sd}:~& [2,1]  \rightarrow & (4,1)~+~(2,2)~+~(1,1)
\nonumber \\
& [2,1]  \rightarrow & (3,0)~+~(0,3)
~~~.
\label{eq-10}
\eeqa
This has to be coupled to $(3,0)$ of the pf-shell:

\beqa
(4,1) \otimes (3,0) & = & (7,1)~+~(5,2)~+~(6,0)
~+~(3,3)
\nonumber \\
&&
~+~(1,4)~+~(4,1)~+~(2,2)
\nonumber \\
(2,2) \otimes (3,0) & = & (5,2)~+~(3,3)~+~(1,4)
~+~(4,1)
\nonumber \\
&&
~+~(2,2)~+~(0,3)~+~(3,0)~+~(1,1)
\nonumber \\
(1,1) \otimes (3,0) & = & (4,1)~+~(2,2)~+~(3,0)~+~(1,1)
\nonumber \\
(3,0) \otimes (3,0) & = & (0,3)~+~(2,2)~+~(4,1)~+~(6,0)
\nonumber \\
(0,3) \otimes (3,0) & = & (0,0)~+~(1,1)~+~(2,2)~+~(3,3)
~~~.
\label{eq-11}
\eeqa

\subsubsection*{Case b)}
\label{case-b-2}

The p-$SU(3)$ irrep is the same. The additional sd-$U(6)$ 
irreps are now $[3,2]$, $[3,1^2]$, $[2^2,1]$, $[2,1^3]$ 
and $[1^5]$. 
The reduction to $SU(3)$ yields

\beqa
\left[3,2\right] & \rightarrow & (6,2)~+~(4,3)~+~(5,1)~+~
(2,4)~+~(3,2)
\nonumber \\
&& ~+~(4,0)~+~(1,3)~+~(2,1)~+~(0,2)
\nonumber \\
\left[3,1^2\right] & \rightarrow & 
(7,0)~+~(4,3)~+~(5,1)~+~(3,2)
\nonumber \\
&& ~+~(0,5)~+~(1,3)~+~(2,1)~+~(1,0)
\nonumber \\
\left[2^2,1\right] & \rightarrow & (5,1)~+~(2,4)~+~(3,2)
~+~(4,0)~+~(1,3)
\nonumber \\
&& ~+~(2,1)~+~(0,2)
\nonumber \\
\left[2,1^3\right] & \rightarrow & (3,2)~+~(1,3)~+~(2,1)
~+~(1,0)
\nonumber \\
\left[ 1^5 \right] & \rightarrow & (0,2)
~~~.
\label{eq-12}
\eeqa

These irreps have to be multiplied by the $(0,1)$
from the p-shell:

\beqa
(6,2) \otimes (0,1) & \rightarrow & 
(6,3)~+~(7,1)~+~(5,2)
\nonumber \\
{\rm twice}~(4,3) \otimes (0,1) & \rightarrow & 
2\{(4,4)~+~(5,2)~+~(3,3)\}
\nonumber \\
{\rm three}~{\rm times}~(5,1) \otimes (0,1) & \rightarrow & 
3\left\{(5,2)~+~(6,0)~+~(4,1)\right\}
\nonumber \\
{\rm twice}~(2,4) \otimes (0,1) & \rightarrow & 
2\left\{(2,5)~+~(3,3)~+~(1,4)\right\}
\nonumber \\
{\rm four}~{\rm times}~(3,2) \otimes (0,1) & \rightarrow & 
4\left\{(3,3)~+~(4,1)~+~(2,2)\right\}
\nonumber \\
{\rm twice}~(4,0) \otimes (0,1) & \rightarrow & 
2\left\{(4,1)~+~(3,0)\right\}
\nonumber \\
{\rm four}~{\rm times}~(1,3) \otimes (0,1) & \rightarrow & 
4\left\{(1,4)~+~(2,2)~+~(0,3)\right\}
\nonumber \\
{\rm four}~{\rm times}~(2,1) \otimes (0,1) & \rightarrow & 
4\left\{(2,2)~+~(3,0)~+~(1,1)\right\}
\nonumber \\
{\rm three}~{\rm times}~(0,2) \otimes (0,1) & \rightarrow & 
3\left\{(0,3)~+~(1,1)\right\}
\nonumber \\
(7,0) \otimes (0,1) & \rightarrow & 
(7,1)~+~(6,0)
\nonumber \\
(0,5) \otimes (0,1) & \rightarrow & 
(0,6)~+~(1,4)
\nonumber \\
{\rm twice}~(1,0) \otimes (0,1) & \rightarrow & 
2\left\{(1,1)~+~(0,0)\right\}
~~~.
\label{eq-12a}
\eeqa

\subsubsection*{Total additional space}
\label{total-2}

Joining the list of irreps of (\ref{eq-11}) and 
(\ref{eq-12a}), which correspond to higher
lying states of supermultiplets, we arrive at

\beqa
&
(4,1)^{13}~+~(6,0)^6~+~(5,2)^8
&
\nonumber \\
&
~+~(3,0)^8+~~(3,3)^{11}~+~(2,2)^{17}~+~(1,1)^{12}
&
\nonumber \\
&
+(0,0)^3~+~(6,3)~+~(7,1)^3~+~(4,4)^2~+~(2,5)^2
&
\nonumber \\
&
~+~(1,4)^9~+~(0,3)^9~+~(0,6)
&
\label{eq-12b}
\eeqa

\subsubsection*{Removing the center of mass motion}
\label{cm-2}

The additional $SU(3)$ irreps, which contain contributions
to the center of mass motion, are obtained multiplying
the list in (\ref{eq-9}) by $(1,0)$:

\beqa
(6,1) \otimes (1,0) & \rightarrow & (7,1)~+~(5,2)~+(6,0)
\nonumber \\
(5,0) \otimes (1,0) & \rightarrow & (6,0)~+~(4,1)
\nonumber \\
{\rm twice}~(4,2) \otimes (1,0) & \rightarrow & 
2\left\{(5,2)~+~(3,3)~+~(4,1)\right\}
\nonumber \\
{\rm twice}~(2,3) \otimes (1,0) & \rightarrow & 
2 \left\{(3,3)~+~(1,4)~+~(2,2)\right\}]
\nonumber \\
{\rm three}~{\rm times}~(3,1) \otimes (1,0) & \rightarrow & 
3\left\{(4,1)~+~(2,2)~+~(3,0)\right\}
\nonumber \\
{\rm three}~{\rm times}~(1,2) \otimes (1,0) & \rightarrow & 
3\left\{(2,2)~+~(0,3)~+~(1,1)\right\}
\nonumber \\
{\rm twice}~(2,0) \otimes (1,0) & \rightarrow & 
2\left\{(3,0)~+~(1,1)\right\}
\nonumber \\
(0,4) \otimes (1,0) & \rightarrow & 
(1,4)~+~(0,3)
\nonumber \\
(0,1) \otimes (1,0) & \rightarrow & (1,1)~+~(0,0)
~~~.
\label{eq-13}
\eeqa

This leads to the list of irreps, containing center of mass
excitations, which have to be removed from the list in
(\ref{eq-12b}).

\beqa
&
(7,1)~+~(5,2)^3~+~(6,0)^2~+~(3,3)^4~+~(4,1)^6
~+~(1,4)^3~+~(2,2)^8
&
\nonumber \\
&
+(3,0)^5~+~(1,1)^6~+~(0,3)^4~+~(0,0)
~~~.
\label{eq-14}
\eeqa

\subsection*{Total additional shell model irreps}
\label{total-3}

Subtracting (\ref{eq-14}) from (\ref{eq-12b}), leads to the
additional appearing $SU(3)$ shell model states

\beqa
&
(4,1)^{7}~+~(6,0)^4~+~(5,2)^5
&
\nonumber \\
&
~+~(3,0)^3+~~(3,3)^7~+~(2,2)^{9}~+~(1,1)^6
&
\nonumber \\
&
+ (0,0)^2~+~(6,3)~+~(7,1)^2~+~(4,4)^2~+~(2,5)^2
&
\nonumber \\
&
~+~(1,4)^6~+~(0,3)^5~+~(0,6)
&
\label{eq-15}
\eeqa

Again, there is no $(10,4)$, $(8,5)$, $(9,3)$, $(7,4)$ 
neither $(5,5)$ (see (\ref{eq-8})) appearing in 
Kat${\bar {\rm o}}$'s list 
\cite{kato}.

\section*{Acknowledgments}
We acknowledge financial support form DGAPA-PAPIIT (IN100421)

\end{document}